# Specificity-determining DNA triplet code for positioning of human pre-initiation complex

Matan Goldshtein* and David B. Lukatsky[†]

*Avram and Stella Goldstein-Goren Department of Biotechnology Engineering, Ben-Gurion University of the Negev, Beer-Sheva, Israel; [†] Department of Chemistery, Ben-Gurion University of the Negev, Beer-Sheva, Israel;

ABSTRACT   The notion that transcription factors bind DNA only through specific, consensus binding sites has been recently questioned. In a pioneering study by Pugh and Venters no specific consensus motif for the positioning of the human pre-initiation complex (PIC) has been identified. Here, we reveal that nonconsensus, statistical, DNA triplet code provides specificity for the positioning of the human PIC. In particular, we reveal a highly non-random, statistical pattern of repetitive nucleotide triplets that correlates with the genome-wide binding preferences of PIC measured by Chip-exo. We analyze the triplet enrichment and depletion near the transcription start site (TSS) and identify triplets that have the strongest effect on PIC-DNA nonconsensus binding. Our results constitute a proof-of-concept for a new design principle for protein-DNA recognition in the human genome, which can lead to a better mechanistic understanding of transcriptional regulation.

Transcription factors (TFs) are proteins that regulate gene expression. An established paradigm that TFs specifically recognize only relatively short (4-20 bp) consensus DNA motifs (1, 2), has been recently challenged by different high-throughput methods both *in vivo* and *in vitro* (3–6). Human pre-initiation complex (PIC) represents one of the most striking examples where design principles of specific protein-DNA recognition remain unknown (5). In particular in a recent study by Pugh and Venters, using the Chip-exo method, no specificity-determining consensus motifs for the positioning of PIC have been identified, thus challenging an established paradigm that the consensus TATA box motif provides the specificity (5).

Here, we reveal that the enrichment level of certain repetitive nucleotide triplets correlate with the genome-wide binding preferences of TFIIB – a key component of PIC (5). Previously, we suggested a model for yeast PIC positioning based on statistical, nonconsensus protein-DNA binding mechanism (6–8). The nonconsensus mechanism predicts that enrichment of certain repetitive DNA sequence elements can lead to an enhanced protein-DNA binding (6–8). Here, we show that this mechanism (albeit with entirely different DNA sequence symmetries) also describes the positioning of the human PIC, using a simple random-binder model based on a 64-letter triplet alphabet, with the human genomic DNA sequence constituting the only input into the model (see below).

In particular, we analyzed the measured genome-wide occupancy of TFIIB (Fig. 1), and revealed that the peak of this occupancy (positioned ~50 bp downstream of TSS, Fig. 1) is characterized by a highly non-random probability distribution of repetitive nucleotide triplets (Fig. 2). This finding has led us to develop a minimal random-binder model based on 64-letter triplet code as follows. We consider a model TF formimg $M$ contacts with DNA, sliding along the DNA sliding window with the width $L$ (Fig. S1). Such sliding window can be positioned at any genomic position. In order to assign the nonconsensus free energy to the middle of the sliding window, we define the partition function

$$Z = \sum_{i=1}^{L-M+1} \exp\left(-U(i)/k_B T\right) \quad (1)$$

where $k_B$ is the Boltzmann constant and $T$ is the temperature, with the interaction potential $U$,

$$U(i) = \sum_{j=i}^{i+M-1} \sum_{\alpha} K_\alpha S_\alpha(j) \quad (2)$$

where each sequence position $i$ corresponds to a DNA triplet, and there are overall 64 possible nucleotide triplets, $\alpha$ (Fig. S1). Here, $K_\alpha$, is the vector containing 64 random energy parameters taken from the Gaussian distribution with the zero mean (for simplicity) and the standard deviation, $\sigma=2k_B T$; and $S_\alpha(j)$ is also a vector of length 64 with all but one zero elements. The only non-zero element (equal to one) of $S_\alpha(j)$ corresponds to the nucleotide triplet of type $\alpha$ located at the sequence position $j$. After generating 250 random TFs, and averaging the resulting free energy,

$$F = -k_B T \ln(Z) \quad (3)$$

with respect to all TFs, we obtain the average nonconsensus free energy for a given genomic position. Moving the sliding window along the genome, and repeating the procedure described above, we obtain the genome-wide average nonconsensus free energy landscape (Fig. 1). This landscape demonstrates a statistically significant, negative correlation with the measured TFIIB binding preferences (inset in Fig. 1). The lower the nonconsensus free energy, the higher the measured TFIIB binding intensity. We have verified that the obtained results are similar for all three possible reading frames (Fig. S2).

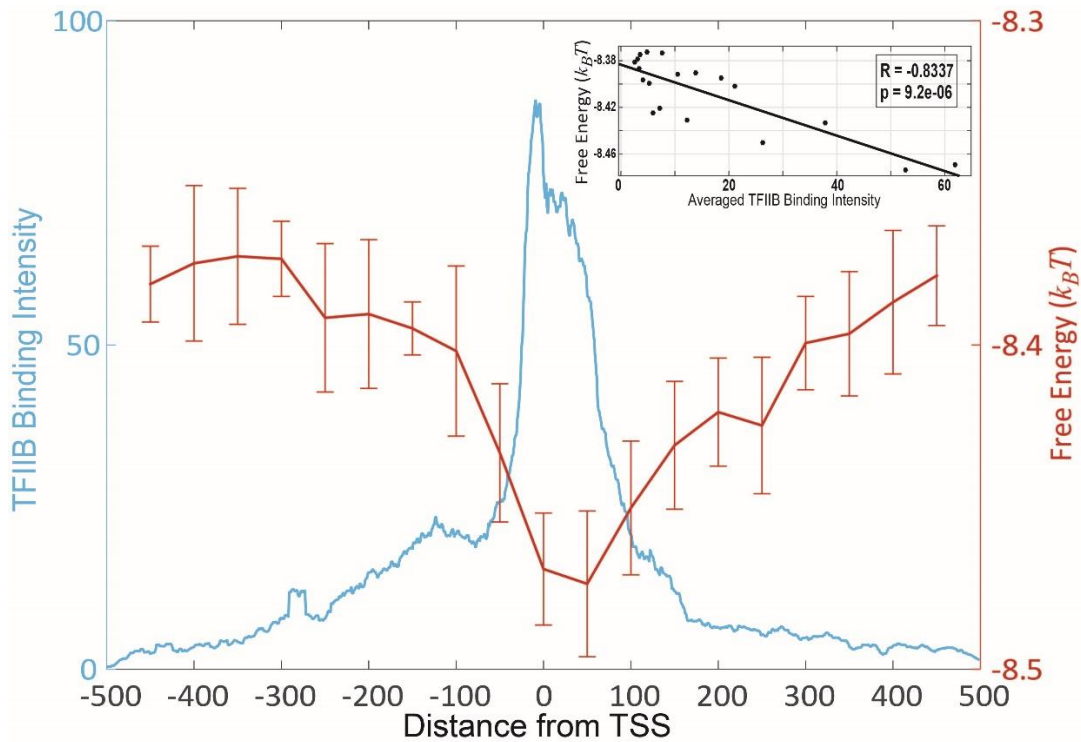

**FIGURE 1 Free energy of nonconsensus triplets based TFIIB-DNA binding negatively correlates with the TFIIB binding intensity.** The computed average free energy of nonconsensus TFIIB-DNA binding and the profile of the average TFIIB binding intensity measured by Pugh and Venters (5) around the TSSs of 8364 genes. The average free energy was calculated every 50 bp, within the interval (-450 bp; 450 bp). In order to compute the free energy, we used a sliding window of 100 bp. To compute error bars, we calculated the mean free energy for each chromosome and divided the results into five randomly chosen subgroups and computed the mean for each subgroup. The error bars are defined as one standard deviation of mean free energy between the subgroups. (Inset) The correlation between the free energy and the TFIIB binding intensity with the Pearson correlation coefficient and the *p*-value.

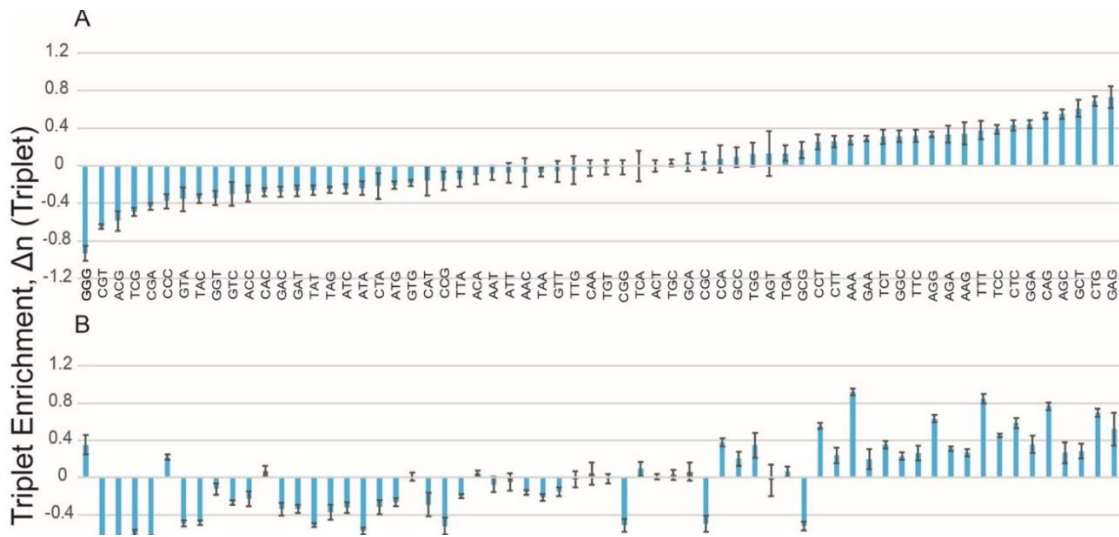

**FIGURE 2 Enrichment levels of 64 nucleotide triplets computed for the genomic regions characterized by high and low TFIIB binding intensity, respectively.** (A) Triplet enrichment in the region of high TFIIB binding intensity, (0 bp; 100 bp). (B) Triplet enrichment in the region of low TFIIB binding intensity, (-450 bp; -350 bp). The enrichment is defined as, $\Delta n = n - \langle n \rangle_{rand}$, where $n$ and $\langle n \rangle_{rand}$ represent the computed average number of nucleotide triplets in the set of actual and randomized DNA sequences, respectively. We used ten randomized DNA replicas in order to compute $\langle n \rangle_{rand}$. Gray colored bars represent triplets that did not exhibit a significant difference based on the two-sample Kolmogorov–Smirnov *p*-value (Table S1). To compute error bars, we divided DNA sequences into four randomly chosen subgroups and computed the mean value of the enrichment for each subgroup. The error bars are defined as two standard deviation of the mean between the subgroups.

Highly non-random distribution of repetitive nucleotide triplets along the human genomic DNA provides the reason for the observed effect (Fig. 2). In particular, we analyzed the enrichment level for 64 possible nucleotide triplets in the region of the highest TFIIB binding intensity positioned in the interval (0;100), and compared this enrichment with the one observed in the interval distant from TSS, (-450;-350) (Fig. 2). The computed triplet enrichment, $\Delta n=n-<n>_{rand}$, is normalized by the GC content in each genomic region separately, and it thus represents a robust measure characterizing the enrichment of repetitive nucleotide triplet patterns. Here, $n$ and $<n>_{rand}$ represent the computed average number of nucleotide triplets in the set of actual and randomized DNA sequences, respectively. We used ten randomized DNA replicas in order to compute $<n>_{rand}$.

In order to further validate statistical significance of our results, we computed the Kolmogorov-Smirnov (KS) $p$-value for each nucleotide triplet (Table S1). This $p$-value provides a statistical significance of the difference between the actual and randomized probability distributions, $P(n)$ and $P(n_{rand})$, respectively (Table S1). For the genomic interval (0;100), the majority (60 out of 64) of computed $p$-values are highly significant (Fig. 2A and Table S1). For example, the enrichment of GAG triplet and the depletion of GGG triplet, provide the strongest signature for the enhanced TFIIB binding intensity (Fig. 2A). The pattern of nucleotide triplet enrichment is entirely different for the interval (-350;-450), with 54 out of 64 computed $p$-values being significant (Fig. 2B and Table S1).

The obtained pattern of nucleotide triplet enrichment (Fig. 2) is validated by the computed pair correlation function, $\eta_{\alpha\alpha}(x)$, representing the probability to find two nucleotides of type $\alpha$ separated by the relative distance, $x$ (Fig. 3). Taken together, our results indicate that the nonconsensus mechanism provides the DNA binding specificity for TFIIB, meaning that the entire distribution of enrichment/depletion levels for the majority of nucleotide triplets (and not just one or two specific triplets) influence the TFIIB binding intensity.

In summary, using statistical mechanics model without any fitting parameters with genomic DNA sequence constituting the only input, we reveal that the nonconsensus nucleotide triplet code constitutes a key signature providing PIC binding specificity in the human genome. Our results need to be further validated in the future using direct *in vitro* methods for measuring TFIIB-DNA binding preferences. Such measurements, using purified proteins and DNA, will clarify the question of how much indirect protein-DNA and nucleosome binding influence our model predictions.

**Author Contributions**

M.G. and D.B.L. designed research, performed research, and wrote the paper.

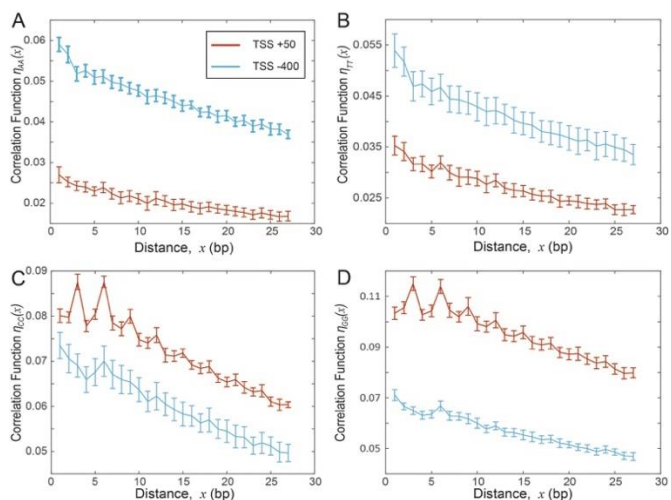

**FIGURE 3 Normalized pair (binary) correlation functions for the nucleotide spacial distribution.** The computed correlation function $\eta_{aa}(x) = (N_{aa}(x)-<N_{aa}(x)>_{rand})/L_0$, where $N_{aa}(x)$ represents the average number of nucleotide pairs of type $a$ separated by the relative distance $x$ bp, and $L_0$ is the width of the window. We used $L_0=100$ bp. We used DNA sequences of 8364 genes for two genomic regions: the region of high TFIIB binding intensity, (0 bp; 100 bp) (red lines); and the region of low TFIIB binding intensity (-450 bp; -350 bp) (blue lines). To compute error bars, we calculated the mean for each chromosome and divided the results into five randomly chosen subgroups and computed the mean for each subgroup. The error bars are defined as one standard deviation of the mean between the subgroups.

1. Berg, O.G., and P.H. von Hippel. 1987. Selection of DNA binding sites by regulatory proteins. J. Mol. Biol. 193: 723–743.
2. Stormo, G.D., and D.S. Fields. 1998. Specificity, free energy and information content in protein-DNA interactions. Trends Biochem. Sci. 23: 109–113.
3. Fordyce, P.M., D. Gerber, D. Tran, J. Zheng, H. Li, J.L. DeRisi, and S.R. Quake. 2010. De novo identification and biophysical characterization of transcription-factor binding sites with microfluidic affinity analysis. Nat. Biotechnol. 28: 970–5.
4. Gordan, R., N. Shen, I. Dror, T. Zhou, J. Horton, R. Rohs, and M.L. Bulyk. 2013. Genomic Regions Flanking E-Box Binding Sites Influence DNA Binding Specificity of bHLH Transcription Factors through DNA Shape. Cell Rep. 3: 1093–1104.
5. Pugh, B.F., and B.J. Venters. 2016. Genomic organization of human transcription initiation complexes. PLoS One. 11.
6. Afek, A., A. Afek, J.L. Schipper, J. Horton, R. Gordân, and D.B. Lukatsky. 2014. Protein-DNA binding in the absence of specific base-pair recognition. Proc. Natl. Acad. Sci. U. S. A. 111: 17140–5.
7. Sela, I., and D.B. Lukatsky. 2011. DNA Sequence correlations shape nonspecific transcription factor-DNA binding affinity. Biophys. J. 101: 160–166.
8. Afek, A., and D.B. Lukatsky. 2013. Genome-wide organization of eukaryotic preinitiation complex is influenced by nonconsensus protein-DNA binding. Biophys. J. 104: 1107–1115.

# Supplemental Information

# Specificity-determining DNA triplet code for positioning of human pre-initiation complex

Matan Goldshtein* and David B. Lukatsky[†]

*Avram and Stella Goldstein-Goren Department of Biotechnology Engineering, Ben-Gurion University of the Negev, Beer-Sheva, Israel; [†] Department of Chemistery, Ben-Gurion University of the Negev, Beer-Sheva, Israel;

# Supporting Material

**Supporting Figures**

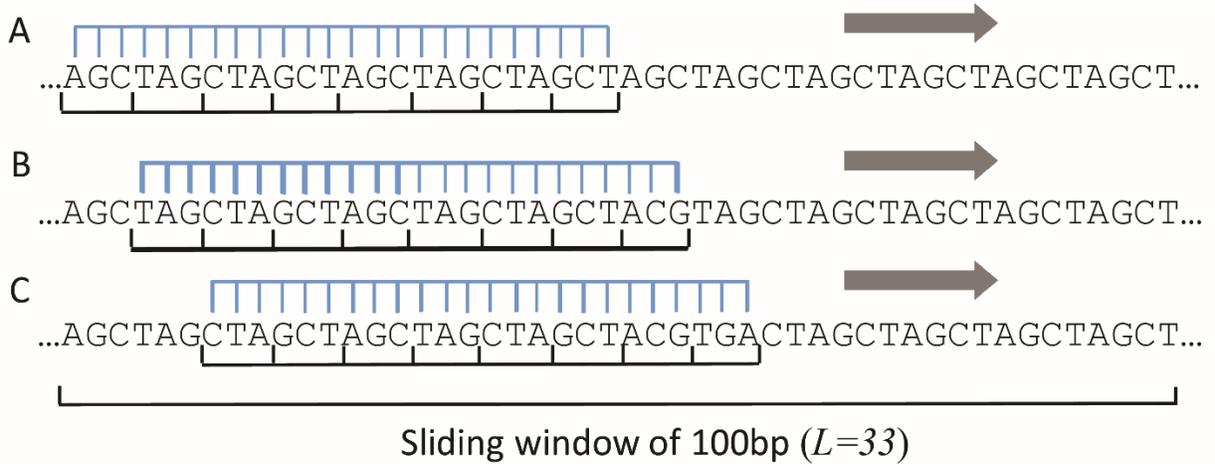

**FIGURE S1 Cartoon illustrating the calculation of the nonconsensus protein-DNA binding energies, *U*, as a model random binder slides along the sliding window**. The interaction contacts of a model protein TF with all DNA nucleotide bases are depicted in blue. The corresponding nucleotide triplets are depicted in black below the DNA strand. In our model we used TF that forms 24 contacts with nucleotide bases (blue), which corresponds to $M=8$ contacts with nucleotide triplets (black). Each model TF slides (gray arrow) along the DNA sequence by 3 bp steps. We used the sliding window with the width 100 bp, which corresponds to $L=33$ nucleotide triplets. The following three examples illustrate the energy calculation as TF slides three consecutive steps along the sliding window: (A) $U(1)=2K_{AGC}+2K_{TAG}+2K_{CTA}+2K_{GCT}$; (B) $U(2)=2K_{TAG}+2K_{CTA}+2K_{GCT}+K_{AGC}+K_{ACG}$; (C) $U(3)=2K_{CTA}+2K_{GCT}+K_{AGC}+K_{ACG}+K_{TAG}+K_{TGA}$. The 64 random energy parameters $K_\alpha$ are drawn from the Gaussian distribution with the zero mean and the standard deviation $\sigma=2k_BT$. These parameters uniquely define a given random binder. In all our calculations we used the free energy averaged over 250 random binders. Therefore, for each DNA sliding window, the procedure described above was repeated for all 250 random binders, each characterized by a different set of $K_\alpha$.

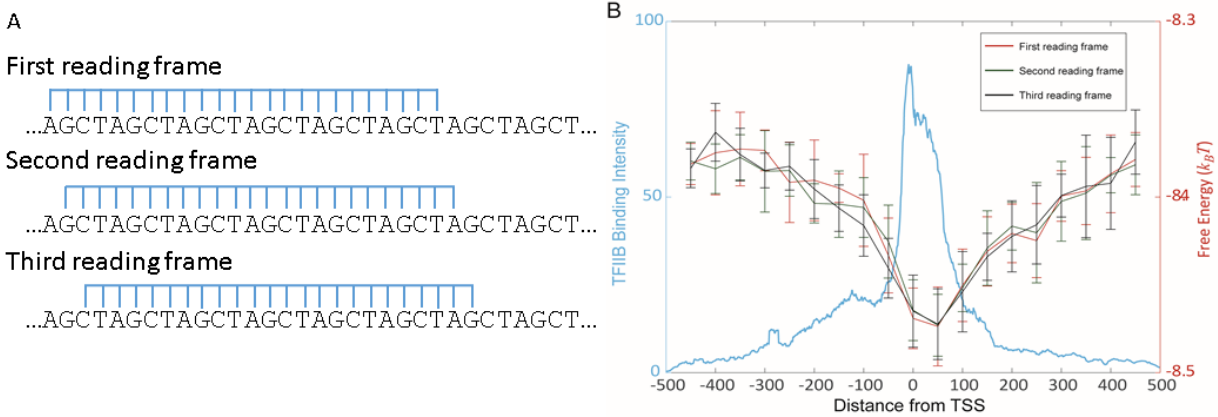

**FIGURE S2 Robustness of the nonconsensus protein-DNA binding free energy landscape computed for different DNA reading frames**. This figure is complementary to Fig. 1 of the main text, and all the definitions and the axes labels are identical to those defined in Fig. 1. (A) Three possible DNA reading frames for a sliding random binder are illustrated. (B) The average free energy of nonconsensus TFIIB-DNA binding for all three possible DNA reading frames, and the measured profile of average TFIIB occupancy around the TSSs of 8364 genes. For each reading frame, the average free energy was calculated every 50 bp, within the interval (-450 bp; 450 bp). The rest of the parameters are identical to those defined in Fig. 1 of the main text.

**Supporting Tables**

|     | Mean [0;100] | Mean [0;100] Rand | KS-test | P-Value | Δ (Mean [0;100]-Mean [0;100] rand) | Mean [-450;-350] | Mean [-450;-350] Rand | KS-test | P-Value | Δ(Mean [-450;-350]-Mean[-450;-350] rand) |
| --- | --- | --- | --- | --- | --- | --- | --- | --- | --- | --- |
| AAA | 0.818 | 0.547 | 1 | 3.194E-44 | 0.272 | 2.607 | 1.689 | 1 | 1.06E-66 | 0.918 |
| AAC | 0.670 | 0.743 | 1 | 5.95E-11 | -0.072 | 1.258 | 1.417 | 1 | 1.46E-14 | -0.159 |
| AAG | 1.292 | 0.951 | 1 | 2.048E-65 | 0.341 | 1.744 | 1.476 | 1 | 1.55E-30 | 0.268 |
| AAT | 0.439 | 0.516 | 1 | 3.259E-12 | -0.077 | 1.357 | 1.431 | 1 | 0.01854 | -0.074 |
| ACA | 0.642 | 0.743 | 1 | 3.675E-23 | -0.102 | 1.469 | 1.423 | 0 | 0.387044 | 0.046 |
| ACC | 1.062 | 1.358 | 1 | 3.673E-57 | -0.296 | 1.379 | 1.610 | 1 | 7.52E-30 | -0.231 |
| ACG | 0.903 | 1.490 | 1 | 1.66E-268 | -0.588 | 0.768 | 1.515 | 1 | 0 | -0.747 |
| ACT | 0.822 | 0.825 | 0 | 0.4344062 | -0.003 | 1.325 | 1.315 | 0 | 0.291719 | 0.010 |
| AGA | 1.287 | 0.952 | 1 | 1.194E-64 | 0.334 | 1.789 | 1.478 | 1 | 7.42E-36 | 0.312 |
| AGC | 2.165 | 1.618 | 1 | 2.61E-123 | 0.547 | 1.793 | 1.524 | 1 | 2.01E-32 | 0.269 |
| AGG | 2.311 | 1.979 | 1 | 9.101E-42 | 0.333 | 2.260 | 1.626 | 1 | 3.3E-122 | 0.634 |
| AGT | 1.100 | 0.970 | 1 | 2.385E-12 | 0.131 | 1.252 | 1.281 | 0 | 0.502634 | -0.029 |
| ATA | 0.258 | 0.494 | 1 | 9.72E-125 | -0.236 | 0.860 | 1.433 | 1 | 4.4E-193 | -0.574 |
| ATC | 0.557 | 0.803 | 1 | 1.631E-97 | -0.246 | 0.994 | 1.314 | 1 | 9.88E-59 | -0.320 |
| ATG | 0.728 | 0.936 | 1 | 1.745E-39 | -0.207 | 1.021 | 1.287 | 1 | 1.49E-57 | -0.266 |
| ATT | 0.516 | 0.590 | 1 | 6.57E-14 | -0.073 | 1.332 | 1.378 | 0 | 0.231387 | -0.047 |
| CAA | 0.730 | 0.755 | 1 | 4.105E-05 | -0.026 | 1.470 | 1.427 | 0 | 0.118211 | 0.044 |
| CAC | 1.074 | 1.356 | 1 | 1.165E-59 | -0.282 | 1.676 | 1.611 | 0 | 0.059437 | 0.065 |
| CAG | 2.155 | 1.623 | 1 | 8.87E-117 | 0.532 | 2.284 | 1.518 | 1 | 2.1E-226 | 0.766 |
| CAT | 0.651 | 0.813 | 1 | 2.143E-36 | -0.162 | 1.024 | 1.314 | 1 | 1.38E-61 | -0.290 |
| CCA | 1.473 | 1.401 | 1 | 0.0024857 | 0.071 | 1.995 | 1.616 | 1 | 3.09E-62 | 0.379 |
| CCC | 2.778 | 3.157 | 1 | 1.899E-12 | -0.378 | 2.719 | 2.497 | 1 | 3.48E-21 | 0.222 |
| CCG | 3.131 | 3.293 | 1 | 5.144E-16 | -0.162 | 1.618 | 2.138 | 1 | 1.1E-96 | -0.520 |
| CCT | 1.976 | 1.722 | 1 | 1.337E-24 | 0.254 | 2.176 | 1.622 | 1 | 7.19E-91 | 0.554 |
| CGA | 1.078 | 1.510 | 1 | 1.27E-139 | -0.432 | 0.827 | 1.513 | 1 | 5.1E-285 | -0.687 |
| CGC | 3.362 | 3.311 | 1 | 3.612E-09 | 0.051 | 1.641 | 2.139 | 1 | 1.68E-91 | -0.498 |
| CGG | 3.790 | 3.808 | 1 | 5.945E-11 | -0.018 | 1.558 | 2.070 | 1 | 3.3E-101 | -0.512 |
| CGT | 1.088 | 1.736 | 1 | 9.85E-294 | -0.647 | 0.708 | 1.445 | 1 | 0 | -0.737 |
| CTA | 0.580 | 0.798 | 1 | 4.591E-72 | -0.218 | 0.992 | 1.309 | 1 | 2.15E-59 | -0.318 |
| CTC | 2.177 | 1.746 | 1 | 2.384E-59 | 0.431 | 2.207 | 1.622 | 1 | 6.22E-94 | 0.585 |
| CTG | 2.579 | 1.890 | 1 | 1.34E-192 | 0.689 | 2.149 | 1.453 | 1 | 5E-199 | 0.697 |
| CTT | 1.331 | 1.074 | 1 | 2.199E-35 | 0.257 | 1.612 | 1.374 | 1 | 5.86E-29 | 0.238 |
| GAA | 1.230 | 0.941 | 1 | 1.228E-54 | 0.289 | 1.671 | 1.473 | 1 | 2.72E-17 | 0.197 |
| GAC | 1.253 | 1.534 | 1 | 1.396E-51 | -0.281 | 1.177 | 1.516 | 1 | 4.38E-71 | -0.340 |
| GAG | 2.750 | 2.021 | 1 | 9.79E-139 | 0.729 | 2.154 | 1.632 | 1 | 1.14E-58 | 0.522 |
| GAT | 0.659 | 0.927 | 1 | 1.163E-84 | -0.268 | 0.943 | 1.279 | 1 | 7.86E-70 | -0.336 |
| GCA | 1.602 | 1.564 | 0 | 0.6949699 | 0.037 | 1.579 | 1.514 | 0 | 0.11215 | 0.064 |
| GCC | 3.419 | 3.327 | 1 | 0.0223591 | 0.092 | 2.345 | 2.142 | 1 | 5.91E-14 | 0.203 |
| GCG | 4.002 | 3.836 | 1 | 4.071E-11 | 0.166 | 1.560 | 2.078 | 1 | 2E-117 | -0.518 |
| GCT | 2.477 | 1.868 | 1 | 1.51E-146 | 0.609 | 1.731 | 1.449 | 1 | 1.67E-32 | 0.283 |
| GGA | 2.427 | 1.987 | 1 | 1.482E-63 | 0.440 | 1.976 | 1.618 | 1 | 9.82E-56 | 0.358 |
| GGC | 4.173 | 3.860 | 1 | 7.107E-15 | 0.314 | 2.307 | 2.079 | 1 | 1.86E-13 | 0.228 |
| GGG | 3.781 | 4.712 | 1 | 2.321E-57 | -0.931 | 2.646 | 2.293 | 1 | 1.25E-35 | 0.352 |
| GGT | 1.729 | 2.073 | 1 | 5.782E-64 | -0.344 | 1.325 | 1.446 | 1 | 9.42E-09 | -0.122 |
| GTA | 0.561 | 0.916 | 1 | 1.87E-163 | -0.354 | 0.789 | 1.281 | 1 | 8.8E-171 | -0.492 |
| GTC | 1.465 | 1.766 | 1 | 5.475E-64 | -0.301 | 1.172 | 1.439 | 1 | 6.5E-44 | -0.267 |
| GTG | 1.912 | 2.097 | 1 | 5.428E-22 | -0.186 | 1.454 | 1.442 | 0 | 0.143382 | 0.012 |
| GTT | 1.093 | 1.153 | 1 | 0.0061886 | -0.060 | 1.132 | 1.284 | 1 | 2.49E-15 | -0.152 |
| TAA | 0.443 | 0.514 | 1 | 4.087E-09 | -0.070 | 1.209 | 1.423 | 1 | 1.38E-21 | -0.214 |
| TAC | 0.436 | 0.786 | 1 | 7.4E-203 | -0.350 | 0.834 | 1.318 | 1 | 1.4E-142 | -0.484 |
| TAG | 0.670 | 0.924 | 1 | 1.065E-76 | -0.254 | 0.911 | 1.284 | 1 | 4.71E-92 | -0.373 |
| TAT | 0.309 | 0.567 | 1 | 6.44E-127 | -0.258 | 0.876 | 1.386 | 1 | 5.8E-150 | -0.509 |
| TCA | 0.830 | 0.834 | 0 | 0.2012681 | -0.004 | 1.409 | 1.315 | 1 | 1.28E-05 | 0.094 |
| TCC | 2.117 | 1.731 | 1 | 8.48E-56 | 0.386 | 2.077 | 1.624 | 1 | 1.74E-71 | 0.453 |
| TCG | 1.254 | 1.743 | 1 | 1.08E-164 | -0.490 | 0.795 | 1.438 | 1 | 5E-259 | -0.644 |
| TCT | 1.402 | 1.094 | 1 | 2.304E-45 | 0.307 | 1.726 | 1.372 | 1 | 2.02E-50 | 0.354 |
| TGA | 1.106 | 0.974 | 1 | 2.102E-13 | 0.133 | 1.355 | 1.290 | 1 | 0.007369 | 0.065 |
| TGC | 1.836 | 1.809 | 0 | 0.2064999 | 0.027 | 1.475 | 1.446 | 0 | 0.184563 | 0.029 |
| TGG | 2.255 | 2.134 | 1 | 4.946E-08 | 0.121 | 1.795 | 1.448 | 1 | 4.98E-51 | 0.347 |
| TGT | 1.138 | 1.156 | 1 | 0.0225222 | -0.018 | 1.262 | 1.274 | 1 | 0.004641 | -0.012 |
| TTA | 0.438 | 0.580 | 1 | 1.268E-36 | -0.143 | 1.187 | 1.387 | 1 | 1.09E-15 | -0.200 |
| TTC | 1.403 | 1.087 | 1 | 4.787E-59 | 0.316 | 1.636 | 1.374 | 1 | 8.9E-32 | 0.262 |
| TTG | 1.104 | 1.151 | 1 | 0.000423 | -0.047 | 1.256 | 1.276 | 0 | 0.848099 | -0.020 |
| TTT | 1.203 | 0.824 | 1 | 4.301E-54 | 0.378 | 2.344 | 1.496 | 1 | 6.85E-71 | 0.848 |

**TABLE S1 This table is complementary to Fig. 2 of the main text. It provides the two-sample Kolmogorov–Smirnov *p*-values for the statistical significance of the enrichment levels for all 64 nucleotide triplets.** Triplets colored in yellow did not show significant enrichment or depletion at [0;100], triplets colored in blue did not show significant enrichment or depletion at [-450;-350], triplets colored in green did not show significant enrichment or depletion at both [0;100] and [-450;-350].